\documentclass[aps,pra,twocolumn]{revtex4}
\usepackage[pctex32]{graphics}

\begin{document}

\title{Dynamical Response of Fermi Condensate to Varying Magnetic Fields}
\author{W. Yi and L.-M. Duan}
\address{FOCUS center and MCTP, Department of Physics,
University of Michigan, Ann Arbor, MI 48109}

\begin{abstract}
We investigate the dynamical response of strongly interacting
ultra-cold fermionic atoms near Feshbach resonance to varying
magnetic fields. Following the experimental practices, we
calculate the response of the atoms to oscillating and to linearly
ramped magnetic fields respectively. For oscillating magnetic
fields, depending on the frequencies and the amplitudes of the
oscillations, the response of the pair excitation gap shows either
linear or rich non-linear behaviour. In addition, both the
spectral studies through the linear response theory and the
time-domain simulations suggest the existence of a resonant
frequency corresponding to the pair dissociation threshold. For
linearly ramped magnetic fields, the response of the excitation
gap shows damped oscillations. The final value of the excitation
gap depends on the rate of the field sweep.
\end{abstract}

\maketitle

\section{Introduction}

Feshbach resonance in ultra-cold Fermi gases, where the energy of
two scattering atoms is magnetically tuned to coincide with that
of a quasi-bound molecular state, has provided a rare opportunity
to study strongly correlated many-body systems with tunable
interactions \cite{1,2}. Recently, considerable experimental
efforts have been dedicated to the
condensation of fermionic atom pairs near Feshbach resonance \cite{3,4,5,6,16}%
. The properties of such a Fermi condensate and its crossover to the
Bose-Einstein condensation (BEC) of diatomic molecules are currently under
active studies \cite{9,8,13,14,12,15}.

In this paper, we investigate the dynamical response of a Fermi
condensate to magnetic field modulations. We study two kinds of
modulations, the sinusoidal oscillation and the linear ramp, both
of which have been frequently applied in experiments
\cite{1,3,4,5,16,26,27}. There have been some recent studies on
the atom-molecule oscillations under a sudden switch of the
magnetic field \cite{14,12}. By considering response of the Fermi
condensate to more general variations of the magnetic field, we
find richer response structures of this strongly interacting
system under different diving frequencies or sweep rates. In
particular, oscillating magnetic fields have been used recently as
important experimental tools to investigate the relaxation time of
the Fermi condensate \cite{16}; while fast magnetic field sweeps
from the BCS side of Feshbach resonance to the BEC side have been
used to probe the properties of the Fermi pairs on the BCS side
\cite{3,4}.

In the case of oscillating magnetic fields, we first develop a
linear response theory in the frequency domain to describe the
dynamics of the Fermi condensate under arbitrary but small
variations of the magnetic field. This spectral analysis is then
compensated by direct time-domain simulations. The result shows
that the response of the Fermi condensate remains linear up to the
pair dissociation frequency, around and above which the
non-linearity of the system becomes dominant, giving rich
non-linear response structure. For linearly ramped magnetic
fields, we perform real time evolutions at different ramping
rates. The responses of the system are essentially damped
atom-molecule oscillations. The final states of the damping
process, however, are dependent on the ramping rates. For
adiabatic ramps, the final states are close to the stationary
state at the final magnetic field of the field sweep. This is
consistent with the experimental practice, where adiabatic sweeps
are applied across the Feshbach resonance from the BCS side
towards the BEC side to efficiently convert Fermions into Feshbach
molecules \cite{26,27}. In contrast, the final state of a sudden
ramp correlates rather with the stationary state at the initial
magnetic field.

In the following, we first derive the equations of motion starting
from the general two-channel Hamiltonian. Our approach is based on
the time-dependent variational theory, by assuming an evolving
pair wavefunction as the ansatz state. Then in Sec. III, we apply
the equations of motion to study the response of the system to
oscillating magnetic fields. We first perform a spectral analysis
using the linear response theory to characterize the response
spectrum, which is later supported by direct time-domain
simulations. In Sec. IV, we apply the equations of motion to study
the response of the system to linearly ramped magnetic fields. We
conclude with a summary in Sec. V.

\section{Model Hamiltonian and The Equations of Motion}

The coupling in strongly interacting Fermi gas near Feshbach
resonance can be described by a two-channel model with the
Hamiltonian \cite{9,7,20,10,12,18}:
\begin{eqnarray}
H &=&\sum_{\mathbf{k},\sigma }\varepsilon _{\mathbf{k}}a_{\mathbf{k},\sigma
}^{\dag }a_{\mathbf{k},\sigma }+\sum_{\mathbf{k}}\left( \varepsilon _{%
\mathbf{k}}/2+\bar{\gamma}\right) b_{\mathbf{k}}^{\dag }b_{\mathbf{k}}
\nonumber \\
&&+\sum_{\mathbf{k},\mathbf{q}}\frac{g}{\sqrt{V}}\left( b_{\mathbf{q}}^{\dag
}a_{\mathbf{k}+\mathbf{q/2},\uparrow }a_{-\mathbf{k}+\mathbf{q/2},\downarrow
}+h.c.\right) ,
\end{eqnarray}
where $\varepsilon _{\mathbf{k}}=\hbar ^{2}\mathbf{k}^{2}/2m$ is the atomic
kinetic energy ($m$ is the mass of the atom), $g$ is the atom-molecule
coupling rate, $V$ is the volume of the system, and $a_{\mathbf{k},\sigma
}^{\dag }$, $b_{\mathbf{k}}^{\dag }$ are the creation operators for the
fermionic atoms and the bosonic bare molecules with momentum $\hbar \mathbf{k%
}$, respectively. Atoms in two different Zeeman levels are labeled
by the spin index $\sigma $( $\sigma =\uparrow ,\downarrow $). The
bare energy detuning $\bar{\gamma}$ of the quasi-bound molecular
level is related to the physical detuning $\gamma _{p}=2\mu
_{B}\left[ B\left( t\right) -B_{0}\right] $ by the standard
renormalization relation $\bar{\gamma}=\gamma
_{p}+g^{2}/V\sum_{\mathbf{k}}\left( 1/2\varepsilon
_{\mathbf{k}}\right) $ \cite{9,8,10}. In the expression for
$\gamma _{p}$, $B\left( t\right) $ is the applied magnetic field
which we take to be time-dependent, and $B_{0}$ denotes the
Feshbach resonance point. In writing this Hamiltonian, we have
ignored the background scattering as it is negligible near
Feshbach resonance.

The basic method to deal with this system at very low temperature
is the crossover theory which is based on the variational
assumption that the ground state of the above Hamiltonian can be
written as a condensate of the dressed molecules \cite{9,8,10,21}.
Here, we extend this variational method to the time-dependent case
(with varying $B\left( t\right) $) by assuming that the state of
the system at any time $t$ can still be written in the form:
\begin{equation}
\left| \Phi \left( t\right) \right\rangle =\mathcal{N}\exp [A\left( t\right)
b_{0}^{\dag }+\sum_{\mathbf{k}}f_{\mathbf{k}}\left( t\right) a_{-\mathbf{k}%
,\downarrow }^{\dagger }a_{\mathbf{k},\uparrow }^{\dagger }]\left|
vac\right\rangle ,
\end{equation}
but with time-varying variational parameters $A\left( t\right) $ and $f_{%
\mathbf{k}}\left( t\right) $ (where $\mathcal{N}$ is the
normalization factor). The parameters $A\left( t\right) $ and
$f_{\mathbf{k}}\left( t\right) $ can be identified with
$\left\langle b_{0}\right\rangle $ and the Cooper-pair
wavefunction in $\mathbf{k}$-space, respectively. This variational
ansatz is reasonable since a condensate state of the form of Eq.
(2) is pretty robust against excitations induced by the magnetic
field variations: it is unlikely to produce single-atom
excitations due to the existence of an energy gap (this is
particularly the case when the rate of the field variation is
smaller than the energy gap); it is also hard to generate
excitations with $\mathbf{q}\neq 0$ from the $\mathbf{q}=0$
condensate due to the momentum conservation.

With the ansatz state above, we minimize the action $S\equiv \int
dt\left\langle \Phi \left( t\right) |\left( i\partial
_{t}-H\right) |\Phi \left( t\right) \right\rangle +h.c.$ under
$\left| \Phi \left( t\right) \right\rangle $. The functional
derivatives of $S$ with respect to the variational parameters
$A^*(t)$ and $f^*_{\mathbf{k}}(t)$ give the dynamical evolution
equations:
\begin{eqnarray}
i\dot{f_{\mathbf{k}}}\left( t\right) &=&2\varepsilon _{\mathbf{k}}f_{\mathbf{%
k}}\left( t\right) -\frac{g}{\sqrt{V}}A^{\ast }\left( t\right) f_{\mathbf{k}%
}^{2}\left( t\right) +\frac{g}{\sqrt{V}}A\left( t\right) , \\
i\dot{A}\left( t\right) &=&\frac{g}{\sqrt{V}}\sum_{\mathbf{k}}\frac{f_{%
\mathbf{k}}(t)}{1+\left| f_{\mathbf{k}}^{2}(t)\right| }+\bar{\gamma}A\left(
t\right) .
\end{eqnarray}

The above evolution equations can be arranged into a more
convenient form if
one introduces the pseudo-spin operators $\sigma _{\mathbf{k}}^{-}=a_{%
\mathbf{k},\uparrow }a_{-\mathbf{k},\downarrow },$ $\sigma _{\mathbf{k}%
}^{+}=a_{-\mathbf{k},\downarrow }^{\dagger }a_{\mathbf{k},\uparrow
}^{\dagger },$ $\sigma _{\mathbf{k}}^{z}=n_{\mathbf{k},\uparrow }+n_{-%
\mathbf{k},\downarrow }-1$ (where $n_{\mathbf{k},\uparrow }=a_{\mathbf{k}%
,\sigma }^{\dag }a_{\mathbf{k},\sigma }$)\cite{11,10,12}. Under the
variational state (2), it is easy to check that the mean values of these
pseudo-spin operators are uniquely related to the pair-wavefunction $f_{%
\mathbf{k}}\left( t\right) $ through $f_{\mathbf{k}}=\left( \left\langle
\sigma _{\mathbf{k}}^{z}\right\rangle +1\right) /\left( 2\left\langle \sigma
_{\mathbf{k}}^{+}\right\rangle \right) $ and the natural constraint $%
4\left\langle \sigma _{\mathbf{k}}^{+}\right\rangle \left\langle \sigma _{%
\mathbf{k}}^{-}\right\rangle +\left\langle \sigma _{\mathbf{k}%
}^{z}\right\rangle ^{2}=1$. With these relations, the evolution
equations (3-4) can be written in terms of the expectation values
of the pseudo-spin operators:
\begin{eqnarray}
i\dot{\left\langle \sigma _{\mathbf{k}}^{-}\right\rangle } &=&2\varepsilon _{%
\mathbf{k}}\left\langle \sigma _{\mathbf{k}}\right\rangle -\Delta
\left\langle \sigma _{\mathbf{k}}^{z}\right\rangle , \\
i\dot{\left\langle \sigma _{\mathbf{k}}^{z}\right\rangle } &=&2\Delta
\left\langle \sigma _{\mathbf{k}}^{-}\right\rangle ^{\ast }-2\Delta ^{\ast
}\left\langle \sigma _{\mathbf{k}}^{-}\right\rangle , \\
i\dot{\Delta} &=&\frac{g^{2}}{V}\sum_{\mathbf{k}}\left\langle \sigma _{%
\mathbf{k}}^{-}\right\rangle +\bar{\gamma}\Delta ,
\end{eqnarray}
where the parameter $\Delta \equiv g\left\langle b_{0}\right\rangle
=gA\left( t\right) $ is physically identified as the excitation gap for
fermionic excitations \cite{21}. The set of equations (5-7) can also be
derived by replacing the operators with their mean values in the Heisenberg
evolution equations for $\sigma _{\mathbf{k}}^{-},\sigma _{\mathbf{k}}^{z}$,
and $b_{0}$ \cite{10,12}. The derivation here provides a different physical
interpretation of the basic approximation: it is equivalent to an extension
of the variational ansatz for the crossover theory \cite{9,8,10,21} to the
time-dependent case.

To investigate how the system responds to the variations of the magnetic
field (which varies the detuning $\bar{\gamma}$), we look at the evolution
of a detectable property, which we choose as the gap function $\Delta \left(
t\right) .$ The gap $\Delta \left( t\right) $ is proportional to the
occupation in the bare molecular level, and is directly measurable, for
instance, through the radio-frequency spectroscopy \cite{22,23} or the bare
molecule population detection \cite{Hulet}. In the subsequent sections, we
will examine the response of the system to magnetic field oscillations and
linear sweeps at different rates, respectively.

\section{Response to Oscillations of the Magnetic Field}

To characterize the response of the system to arbitrary, yet small, magnetic
field oscillations, we expand $\Delta \left( t\right) ,\left\langle \sigma _{%
\mathbf{k}}^{-}\left( t\right) \right\rangle ,\left\langle \sigma _{\mathbf{k%
}}^{z}\left( t\right) \right\rangle $ around their stationary
values, and linearize the resultant evolution equations. The
stationary values are obtained from the conventional crossover
theory: we minimize $\left\langle H-\mu N\right\rangle $ with a
fixed initial detuning $\gamma _{p}=\gamma _{0} $, and get the
following gap and number equations characterizing the stationary
state \cite{9}:
\begin{eqnarray}
\left( \gamma _{0}-2\mu \right)  =\frac{g^{2}}{V}(\sum_{k}\frac{1}{2E_{k}}-%
\frac{1}{2\epsilon _{\mathbf{k}}}), \\
n =2\frac{\Delta _{0}^{2}}{g^{2}}+\sum_{k}\left( 1-\frac{\epsilon _{%
\mathbf{k}}-\mu }{E_{\mathbf{k}}}\right) ,
\end{eqnarray}
where $N=\sum_{\mathbf{k},\sigma }a_{\mathbf{k},\sigma }^{\dag }a_{\mathbf{k}%
,\sigma }+2\sum_{\mathbf{k}}b_{\mathbf{k}}^{\dag }b_{\mathbf{k}}$ denotes
the total atom number, $\mu $ is the Lagrange multiplier with the meaning of
atomic chemical potential, and $E_{\mathbf{k}}=\sqrt{(\varepsilon _{\mathbf{k%
}}-\mu )^{2}+ \Delta _{0}^{2}}$ denotes the single-atom excitation
spectrum in the stationary case. From the stationary equations,
the gap $\Delta _{0}$ and the chemical potential $\mu $ in the
stationary state can be solved, from which the stationary values
of the pseudospins are obtained as follows:
\begin{eqnarray}
\left\langle \sigma _{\mathbf{k}}^{-}\right\rangle  &=&-\frac{\Delta _{0}}{%
2E_{\mathbf{k}}} \\
\left\langle \sigma _{\mathbf{k}}^{z}\right\rangle  &=&-\frac{\epsilon _{%
\mathbf{k}}-\mu }{E_{\mathbf{k}}}.
\end{eqnarray}

From the linearized evolution equations, we can find out the
response relation between $\delta \Delta (t)\equiv \Delta
(t)-\Delta _{0}e^{-2i\mu t}$
and $\delta \gamma \left( t\right) \equiv \gamma _{p}-\gamma _{0}=2\mu _{B}%
\left[ B\left( t\right) -B\left( 0\right) \right] $, where
$B\left( 0\right) $ is the initial magnetic field. It is
convenient to express this response relation in terms of their
Fourier transforms $\delta \Delta (\omega )\equiv \int \delta
\Delta (t)e^{-i\left( \omega -2\mu \right) t}dt$ and $\delta
\gamma (\omega )\equiv \int \delta \gamma (t)e^{-i\omega t}dt$.
The general response relation can then be written as $\delta
\Delta (\omega )=P\left(
\omega \right) \delta \gamma (\omega )$, with the response function $%
P(\omega )$ given by:
\begin{widetext}
\begin{equation}
P(\omega)=\frac{\Delta_0\left(F(\omega)-G^{\ast}(-\omega)+\omega-(\gamma_0-2\mu)\right)}
{(\omega+G(\omega)+(\gamma_0-2\mu))(-\omega+G^{\ast}(-\omega)+(\gamma_0-2\mu))-F(\omega)F^{\ast}(-\omega)}
,
\end{equation}

in which
\begin{displaymath}
F(\omega)=\frac{g^2}{2\pi^2E_F}\int k^2dk
\frac{\Delta_0^2}{4E_k(E_k^2-\frac{\omega^2}{4})},\qquad
G(\omega)=\frac{g^2}{2\pi^2E_F}\int k^2dk
(\frac{-E_k^2-(\varepsilon_k-\mu)^2+(\varepsilon_k-\mu)\omega}{4E_k(E_k^2-\frac{\omega^2}{4})}+\frac{1}{2\varepsilon_k}).
\end{displaymath}
\end{widetext}

The response function $P(\omega )$ specified by the above
equations can not be directly compared with the experimental
measurements as it is $\left| \Delta \right| $ that is measured.
It is easy, however, to construct the response function $R\left(
\omega \right) $ for $\delta |\Delta |\equiv |\Delta |- \Delta
_{0} $ from $P\left( \omega \right) $ with the following simple
connection:
\begin{equation}
R\left( \omega \right) =\frac{1}{2}[P(\omega )+P^{\ast }(-\omega )].
\end{equation}
The resultant spectrum for $R\left( \omega \right) $ is
numerically calculated and shown in Fig. 1, with $\omega $ in the
range $-2 \Delta _{0} <\omega <2 \Delta _{0} $ \cite{note}, and
other parameters comparable with the $^{40}K$ experiments
\cite{3}. The spectrum shows that the response of the Fermi
condensate is linear for small driving fields with the response
function only weakly dependent on the driving frequency. The
spectrum also shows resonance behaviour at $\omega =2\Delta _{0}$,
which can be associated with the pair dissociation frequency.
\begin{figure}[tbp]
\includegraphics{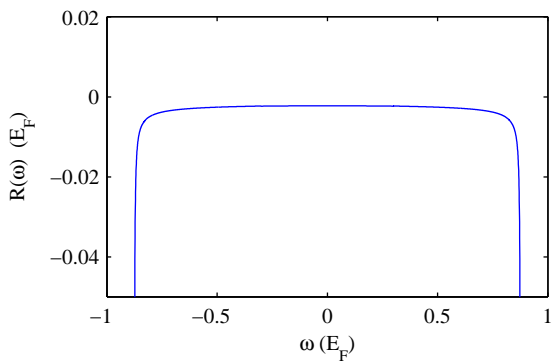}
\caption[Fig.1 ]{(Color online) The response function shown in the range $\left| \protect%
\omega \right| <2 \Delta _{0} $ (here, $2 \Delta
_{0} \sim 0.87E_{F}$, and $\protect\omega $ is in the unit of $E_{F}$%
). Other parameters are comparable with the ${}^{40}K$ experiments
\protect\cite{3}, where $g\protect\sqrt{n}\sim 13.8E_{F}$ ($n$ is
the atom number density and $\protect\gamma _{0}\sim 100E_{F}$).}
\end{figure}

\begin{figure*}[tbp]
\includegraphics{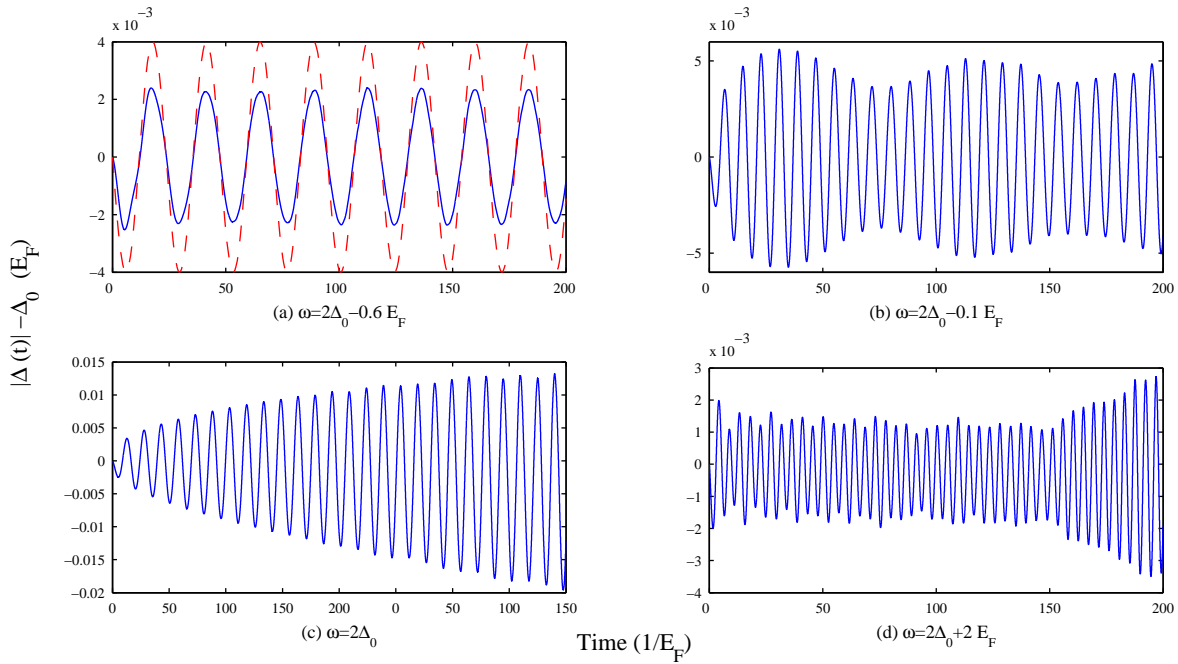}
\caption[Fig.2 ]{(Color online) The real-time (in the unit of
$1/E_{F}$,$\sim 120\mu s$ for ${}^{40}K$ experiments \cite{3})
evolutions of the pair excitation gap $|\Delta (t)|$ (in the unit
of $E_{F}$) shifted by the initial gap $\Delta _{0}$ ($\Delta
_{0}\sim 0.433E_{F}$ ). The system is driven by sinusoidally
oscillating magnetic fields at different driving frequencies
$\protect\omega $. The solid lines represent the evolutions of the
excitation gap $\protect\delta|\Delta (t)|$, while the dashed line
is the re-scaled magnetic field variation $B(0)-B(t)$. In all
cases, $\protect\gamma _{amp}=1E_{F}$, $g\protect\sqrt{n}\sim
13.8E_{F}$, and $\protect\gamma _{0}=100E_{F}$.}
\end{figure*}

To complement the above linear response calculations, we also
simulate the evolutions of the gap function $\Delta \left(
t\right) $ from Eqs. (5-7) directly in the time domain, with
sinusoidal magnetic fields at different driving frequencies. With
this temporal simulation, we would like to first confirm the
resonance predicted by the linear response theory and second find
out the validity region for the linear response theory and reveal
some non-linear behaviours. In the simulation, we start from an
equilibrium state at the positive detuning ($B\left( 0\right)
>B_{0}$), and assume that the variation of the detuning takes the
form $\delta \gamma \left( t\right) =\gamma _{p}-\gamma
_{0}=\gamma _{amp}\sin \left( \omega t\right) $, with $\gamma
_{0}=100E_{F}$ (corresponding to $(B\left( 0\right) -B_{0})\sim 0.2G$ for $%
^{40}K$) and a small oscillation amplitude $\gamma _{amp}=1E_{F}$ ($%
E_{F}=\left( 3\pi ^{2}\hbar ^{3}n\right) ^{2/3}/2m$ is the Fermi
energy, where $n$ is the atom number density and $m$ is the mass
of the atom). The results of the evolutions are shown in Fig. 2
with several characteristic driving frequencies. For driving
frequencies much smaller than $2 \Delta _{0} $ (Fig. 2a), the
$\delta\left|\Delta \left( t\right) \right| $ follows the same
sinusoidal curve as $\left(B\left( 0\right) -B\left(
t\right)\right) $ with almost no detectable delay. This shows that
the system responds very fast to the changes of the magnetic
field, with a time scale much shorter than the thermal relaxation
time reported in the recent experiment \cite{16,note2}. When the
driving frequency approaches $2 \Delta _{0} $ (Fig. 2b), some
non-linearities show up in the response with a periodic modulation
of the oscillation amplitude. For $\omega =2 \Delta _{0} $ (Fig.
2c), the resonance behaviour is apparent. The amplitude of the
oscillation continuously increases within the time interval
attainable in the simulation. For the driving frequency
significantly larger than $2 \Delta _{0} $ (Fig. 2d), the response
is almost linear at the beginning, then the nonlinearity shows up
causing increase of the oscillation amplitude. We have also
simulated the time evolution of the gap function $\Delta \left(
t\right) $ with a much larger oscillation amplitude of the driving
field corresponding to $\gamma _{amp}=50E_{F}$. For small
non-resonant driving frequencies, the response is still linear
even with such a large driving amplitude. For instance, with the
same $\omega $ as in Fig. 2a but $\gamma _{amp}=50E_{F}$, the
response curve of $\delta|\Delta \left( t\right)| $ still has the
same shape, although its amplitude is amplified by a factor of
$50$ as one expects from the linear response theory. For driving
frequencies larger than $2\left| \Delta _{0}\right| $, even if
they are non-resonant, rich non-linear structure shows up in the
response, as one can see in Fig. 3.

\begin{figure}[tbp]
\includegraphics{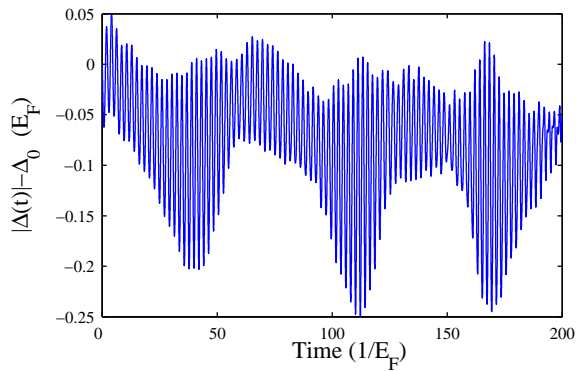}
\caption[Fig.3 ]{(Color online) The real-time evolution of the
pair excitation gap $|\Delta (t)|$ shifted by the initial gap
$\Delta_0$. The same parameters as in Fig. 3d but with a larger
driving amplitude $\protect\gamma _{amp}=50E_{F}$, which roughly
corresponds to $0.15G$ in the ${}^{40}K$ experiments \cite{3}.}
\end{figure}

\section{Response to Linearly Ramped Magnetic Fields}

Another branch of important experimental practice is to probe the
properties of the Fermi condensate by ramping the magnetic field
from the BCS side of the Feshbach resonance to the BEC side. Slow
adiabatic ramps are used to create molecular condensates
\cite{26,27}, while fast sudden ramps are used to project the
correlated pairs onto Feshbach molecules so as to probe the
properties of the Fermi condensate on the BCS side of the
resonance \cite{3,4}. It is desirable then, to understand the
ramping process and the effects of the ramping rate. In this
section, we study the response of the system to linearly ramped
magnetic fields with different ramping rates.

\begin{figure}[tbp]
\includegraphics{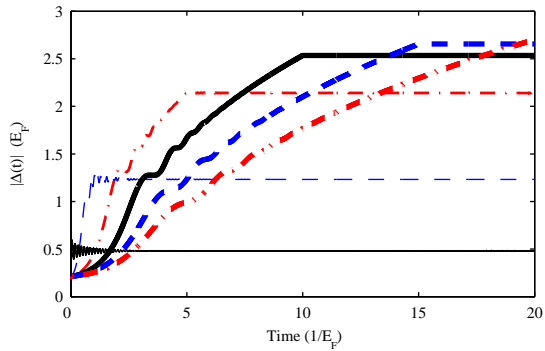}
\caption[Fig.4 ]{(Color online) The real-time evolutions of the pair excitation gap $%
|\Delta(t)|$ under linear magnetic field sweeps with different
sweep rates. While the thin solid curve represents the evolution
under a sudden ramp, the ramping time $T_{ramp}$, as defined in
the text, for the rest of the curves are: $T_{ramp}=1/E_F$ for the
thin dashed curve, $T_{ramp}=5/E_F$ for the
thin dash-dot curve, $T_{ramp}=10/E_F$ for the thick solid curve, $%
T_{ramp}=15/E_F$ for the thick dashed curve, and $T_{ramp}=20/E_F$ for the
thick dash-dot curve. The atom-molecule coupling rate is taken to be $g%
\protect\sqrt{n}\sim 10E_F$, with the initial detuning $\protect\gamma%
_0=100E_F$, and the final detuning $\protect\gamma_f=-500E_F$.}
\end{figure}

\begin{figure}[tbp]
\includegraphics{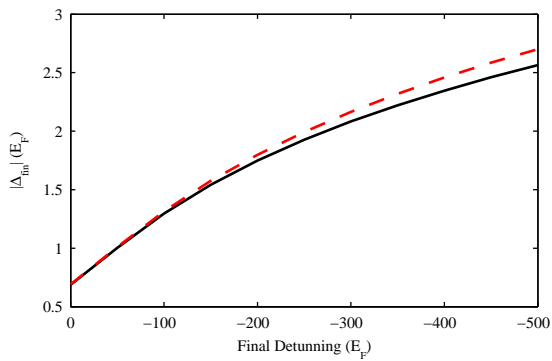}
\caption[Fig.5 ]{(Color online) Comparison between the final
excitation gap values resulting from slow ramps
($T_{ramp}=20/E_F$) to different final detunings, with those of
the stationary states at the corresponding final detunings. The
solid line represents the stationary gap values, and the dashed
line represents the final gap values of the slow ramps. The
starting points of all the field ramps are fixed at
$\protect\gamma_0=100E_F$, while the final detunings are varied
over the range of $\protect\gamma_f=0\sim-500E_F$. The
atom-molecule coupling is taken to be
$g\protect\sqrt{n}\sim10E_F$.}
\end{figure}

\begin{figure}[tbp]
\includegraphics{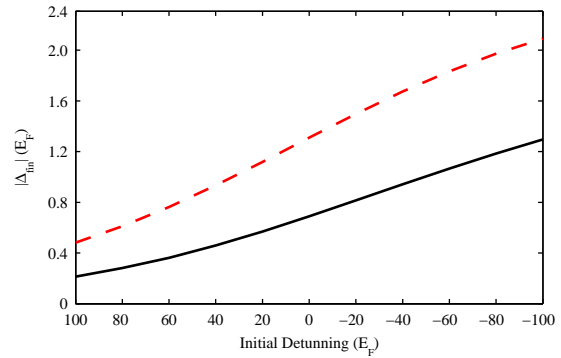}
\caption[Fig.6 ]{(Color online) Comparison between the the final
excitation gap values resulting from sudden ramps from different
initial detunings, with those in the stationary states at the
corresponding initial detunings. The solid line represents the
stationary gap values, and the dashed line corresponds to the
final gap values of the sudden ramps. The final detunings of all
the field ramps are fixed at $\protect\gamma_f=-500E_F$, while the
initial detunings are varied over the range of
$\protect\gamma_0=100E_F\sim-100E_F$. The atom-molecule coupling
is taken to be $g\protect\sqrt{n}\sim 10E_F$.}
\end{figure}

Similar to the previous section, where we vary the magnetic field by
modulating the detuning in Eq. (7), now we impose a detuning in the form $%
\gamma _{p}=\gamma _{0}-(\gamma _{0}-\gamma _{f})t/T_{ramp}$, where $\gamma
_{0},\gamma _{f}$ are the initial and final detunings respectively, and $%
T_{ramp}$ is the duration of the field sweep. To facilitate
comparison between processes with different ramping rates, once
the detuning has been ramped to $\gamma _{f}$, the final detuning,
we let the system evolve under this constant magnetic field.
Again, we use the pair excitation gap $\Delta (t)$ to represent
the response of the Fermi condensate. As the numerical simulation
is quite time consuming at large coupling rates, for the sake of
efficiency, we set the atom-molecule coupling rate to be
$g\sqrt{n}\sim 10E_{F}$, which is slightly smaller than that of
the ${}^{40}K$ system, where $g\sqrt{n}\sim 13.8E_{F}$\cite{3}.
Qualitatively, the results should not be affected by this change
of coupling rate. We perform real-time evolutions of the
excitation gap based on Eqs. (5-7), and the results are shown in
Fig. 4. The most prominent feature of the results is that the
excitation gap undergoes damped oscillations under fast field
sweeps. This result is in agreement with Ref. \cite{14} for the
particular case of a sudden sweep, but is in contrast to the
undamped oscillations reported in \cite{12}. We expect that this
kind of damped oscillations is a general response behaviour for a
system with many degrees of freedom as is the case here; the
undamped oscillations on the other hand should be a more
artificial effect resulting from some unrealistic constraints. For
slower sweeps, the damping also occurs along with the sweep but
with a relatively smaller amplitude; and if the sweep rate is slow
enough, the absolute value of the excitation gap will follow the
field sweep almost adiabatically. Another important feature is
that the final state of the system, reflected in the absolute
value of the final gap, depends on the rate of the field sweep.
Even though the initial and the final detunings are set to be the
same in all the evolutions in Fig. 4, there appear to be
considerable differences in the values of the final gap at
different ramping rates. These rate-dependent ``stationary
states'' are different from the stationary states calculated in
Sec. III with Eqs. (8-9). The final state at a given ramping rate
is one such that the absolute value of the excitation gap is not
evolving but its phase changes with time. Even under such a final
``stationary'' state with a constant $\left| \Delta \left(
t\right) \right| $, numerical analysis shows that there still
appear to be complicated momentum-dependent
oscillations of the pseudospins $\left\langle \sigma _{\mathbf{k}%
}^{+}\right\rangle $, $\left\langle \sigma
_{\mathbf{k}}^{-}\right\rangle $, and $\left\langle \sigma
_{\mathbf{k}}^{z}\right\rangle $, which effectively describe the
shape oscillations of the dressed molecules. The amplitude of such
oscillations decreases as the sweep rate approaches the adiabatic
limit, and in the adiabatic limit, the state just follows the real
stationary state of the system determined by Eqs. (8-9) as the
detuning decreases.

Among these many ramping-rate dependent final states, of
particular interest to us are the final state of a sudden ramp and
the final state of an adiabatic ramp, due to their close
experimental relevance. To study the property of the final states,
we compare the excitation gap in the stationary state at either
the initial or the final detuning with the gap in the final state
of the evolution. One can either fix the initial detuning while
changing the final detuning of the field sweep or the other way
around. We have plotted in Fig. 5 and Fig. 6. the results of our
simulations. For a very slow ramp, if one fixes the initial
detuning and varies the final detuning, the final gap would
approach the stationary values of the gap at those different
$\gamma _{f}$. The differences between the two curves in Fig. 5
can be made smaller by decreasing the ramping rate further. This
is in agreement with the experiment \cite{26}, where slow field
sweeps, which roughly correspond to ramp time of over $100/E_{F}$
to cover the same detuning range in Fig. 4, were applied to create
molecules. In a sudden ramp, one needs to fix the final detuning
while varying the initial starting point. It appears, as in Fig.
6, that there exist correlations between the final gap and the
initial gap, while both are much smaller than the stationary value
of the gap at the final detuning. There are no such correlations
if one fixes the initial detuning and vary the final detuning.
This result is in agreement with Ref. \cite{14} from a single
channel calculation. The finding also agrees with the experimental
practice that a fast projection would allow us to probe
information in the initial state.

\section{Summary}

In summary, we have investigated the dynamical response of a Fermi
condensate near Feshbach resonance to external magnetic field
modulations. For a sinusoidal modulation with small amplitude and
frequency, the response of the Fermi condensate linearly follows
the modulation. The linear behaviour breaks down when the
frequency of the modulated field approaches the pair dissociation
frequency $2\Delta _{0}$. The response of the system will be
composed of highly non-linear structures if the frequency
increases above this threshold. Calculations of the response
spectrum using the linear response theory as well as real time
evolutions consistently support the conclusions. We have also
calculated the real time evolutions of the pair excitation gap
under linearly ramped magnetic fields, the results of which are
consistent with the experimental observations. The general feature
of the response to the field sweep is the damped oscillations,
which are more manifest in faster sweeps. The final state of the
sweep depends on the rate of the sweep. While the final state of a
slow ramp is quite close to the stationary state at the final
detuning; that of a sudden ramp correlates with the stationary
state at the initial detuning, as can be seen from the
correlations between the final gap and the initial gap in a sudden
ramp. The results are in agreement with the experimental practices
that adiabatically slow field sweeps efficiently convert Fermion
pairs on the BCS side of the Feshbach resonance into bound
molecules deep in the BEC side; and that sudden field sweeps
project the Fermion pairs onto the molecules, so that the
properties of the initial state can be probed via measurements of
the final state.

This work was supported by the NSF awards (0431476), the ARDA under ARO
contracts, and the A. P. Sloan Fellowship.


\begin{thebibliography}{99}
\bibitem{note}  As discussed before, the variational ansatz (2) should be a
good approximation for $\omega $ in the range $\left| \omega
\right| <2 \Delta _{0} $.
\bibitem{note2}  When the condensed dressed molecules (with $\mathbf{q}=0$)
and the thermal cloud (with $\mathbf{q\neq }0$) are pushed into a
non-equilibrium state, the thermal relaxation time, which characterizes the
time scale for them to re-obtain the thermal equilibrium, is long compared
with $1/E_{F}$.
\end{thebibliography}
\end{document}